\documentclass[a4paper,pdfLaTeX]{article}
\usepackage{aqis}
\usepackage{bm}
\usepackage{cite}
\usepackage{color}
\usepackage{hyperref}
\hypersetup{colorlinks=true,citecolor=magenta,linkcolor=magenta,urlcolor=magenta}

\begin{document}

\title{
The first-order Trotter decomposition in the dynamical-invariant basis
}

\author{
Takuya Hatomura
  \affiliation{1}
  \email{\href{mailto:takuya.hatomura@ntt.com}{takuya.hatomura@ntt.com}}
  }
  
\address{1}{
NTT Basic Research Laboratories \& NTT Research Center for Theoretical Quantum Physics, \\ NTT Corporation, Kanagawa 243-0198, Japan
}

\abstract{
The Trotter decomposition is a basic approach to Hamiltonian simulation (digital quantum simulation). 
The first-order Trotter decomposition is the simplest one, whose deviations from target dynamics are of the first order of a small coefficient in terms of the infidelity. 
In this paper, we consider the first-order Trotter decomposition in the dynamical-invariant basis. 
By using a state-dependent inequality, we point out that deviations of this decomposition are of the second order of a small coefficient. 
Moreover, we also show that this decomposition includes a useful example, i.e., digital implementation of shortcuts to adiabaticity by counterdiabatic driving. 
}

\keywords{Hamiltonian simulation, Trotter decomposition, dynamical invariant}

\maketitle


\section{Introduction}

Quantum simulation, which is also referred as Hamiltonian simulation, is one of the most promising quantum technologies. 
In quantum simulation (Hamiltonian simulation), we simulate (Hamiltonians of) target quantum systems by using other programable quantum systems~\cite{Feynman1982}. 
Since degrees of freedom in quantum systems increase in an exponential way as their components increase, quantum simulation has clear advantage against classical simulation. \\

The Trotter formulae~\cite{Trotter1959,Suzuki1976} are often used to decompose time-evolution operators of target quantum systems into sequences of simulatable unitary operators~\cite{Lloyd1996}. 
The high-order Trotter formulae~\cite{Suzuki1976} give precise simulation, but the depth of quantum circuits tends to be deep. 
Other approaches to Hamiltonian simulation, e.g., the Taylor series expansion and the linear combination of unitaries~\cite{Berry2015}, quantum signal processing~\cite{Low2017}, etc., have also been proposed for realizing precise simulation with relatively shallow quantum circuits. \\

In this paper, we revisit the simplest matrix exponential formula, i.e., the fist-order Trotter decomposition~\cite{Trotter1959}. 
First, as a preliminary, we introduce the dynamical invariant and the Lewis-Riesenfeld theory~\cite{Lewis1969}. 
Next, we introduce a state-depdendent inequality which enables us to precisely evaluate digitization errors~\cite{Hatomura2022}. 
Then, we consider the first-order Trotter decomposition in the dynamical-invariant basis. 
We will find that dominant errors vanish in this decomposition and the scaling of digitization errors is better than the conventional one which arises from the first-order Trotter decomposition~\cite{Lloyd1996}. 
Finally, we show that this specific decomposition includes a useful example, i.e., digitized counterdiabatic driving~\cite{Hegade2021,Chandarana2022,Hegade2022,Hatomura2023}. \\

\section{Preliminary}

Suppose that a given quantum system is governed by the Schr\"odinger equation
\begin{equation}
i\hbar\frac{\partial}{\partial t}|\Psi(t)\rangle=\hat{H}(t)|\Psi(t)\rangle,
\label{Eq.Seq}
\end{equation} 
where $|\Psi(t)\rangle$ is its dynamics and $\hat{H}(t)$ is its Hamiltonian.

\paragraph{Dynamical invariant.}
A dynamical invariant $\hat{F}(t)$ is an Hermitian operator which satisfies the von Neumann equation
\begin{equation}
i\hbar\frac{\partial}{\partial t}\hat{F}(t)-[\hat{H}(t),\hat{F}(t)]=0. 
\end{equation}
Note that the density operator is a trivial example, but there are infinite dynamical invariants. 
We can easily confirm that eigenvalues of dynamical invariants are independent of time, i.e., their time-dependence comes from their eigenvectors.

In the dynamical-invariant basis $\{|\phi_n(t)\rangle\}$, which is the set of the eigenvectors of a dynamical invariant, off-diagonal elements of the Hamiltonian are given by
\begin{equation}
\langle\phi_m(t)|\hat{H}(t)|\phi_n(t)\rangle=i\hbar\langle\phi_m(t)|\partial_t\phi_n(t)\rangle,\quad\text{for }m\neq n,
\label{Eq.off.hamdy}
\end{equation}
where $|\partial_t\phi_n(t)\rangle=(\partial/\partial t)|\phi_n(t)\rangle$~\cite{Lewis1969}. 
Then, the Hamiltonian can be expressed as
\begin{equation}
\begin{aligned}
\hat{H}(t)=&\sum_n\langle\phi_n(t)|\hat{H}(t)|\phi_n(t)\rangle|\phi_n(t)\rangle\langle\phi_n(t)| \\
&+i\hbar\sum_{\substack{m,n \\ (m\neq n)}}|\phi_m(t)\rangle\langle\phi_m(t)|\partial_t\phi_n(t)\rangle\langle\phi_n(t)|,
\end{aligned}
\label{Eq.ham.divide}
\end{equation}
where the first term is diagonal and the second term is off-diagonal in the dynamical-invariant basis~\cite{Chen2011,Takahashi2017}.

\paragraph{Lewis-Riesenfeld theory~\cite{Lewis1969}.}
By using the dynamical-invariant basis, the solution of the Schr\"odinger equation (\ref{Eq.Seq}) is given by
\begin{equation}
\begin{aligned}
&|\Psi(t)\rangle=\sum_nc_n(0)e^{i\kappa_n(t)}|\phi_n(t)\rangle, \\
&\kappa_n(t)=\frac{1}{\hbar}\int_0^tdt^\prime\langle\phi_n(t^\prime)|\left(i\hbar\frac{\partial}{\partial t^\prime}-\hat{H}(t^\prime)\right)|\phi_n(t^\prime)\rangle, 
\end{aligned}
\label{Eq.LRstate}
\end{equation}
where the coefficient $c_n(0)$ determines the initial state and $\kappa_n(t)$ is the Lewis-Riesenfeld phase. 
We can easily confirm this fact by considering the time derivative of the state (\ref{Eq.LRstate}) and by using Eq.~(\ref{Eq.off.hamdy}).

\section{Results}

\subsection{State-dependent inequality for Hamiltonian simulation}

In this section, we introduce a state-dependent inequality for precisely evaluating digitization errors~\cite{Hatomura2022}. 
For this purpose, we introduce the Fubini-Study angle.

\paragraph{Fubini-Study angle.}
For given two quantum states, $|\psi\rangle$ and $|\phi\rangle$, the Fubini-Study angle (see, e.g., Ref.~\cite{Wootters1981}) is defined by 
\begin{equation}
\mathcal{L}(|\psi\rangle,|\phi\rangle)=\arccos|\langle\psi|\phi\rangle|. 
\end{equation}
The Fubini-Study angle is distance, i.e., it satisfies (i) the identity of indiscernibles, $\mathcal{L}(|\psi\rangle,|\phi\rangle)=0\Leftrightarrow|\psi\rangle=|\phi\rangle$ except for a grobal phase factor; (ii) symmetry, $\mathcal{L}(|\psi\rangle,|\phi\rangle)=\mathcal{L}(|\phi\rangle,|\psi\rangle)$; and (iii) subadditivity, $\mathcal{L}(|\psi\rangle,|\phi\rangle)\le\mathcal{L}(|\psi\rangle,|\chi\rangle)+\mathcal{L}(|\chi\rangle,|\phi\rangle)$ for any quantum state $|\chi\rangle$. 
It also satisfies (iv) unitary invariance, $\mathcal{L}(|\psi\rangle,|\phi\rangle)=\mathcal{L}(\hat{U}|\psi\rangle,\hat{U}|\phi\rangle)$ for any unitary operator $\hat{U}$. \\

Now, we consider the dynamics $|\Psi(t)\rangle$ governed by the Schr\"odinger equation and its digitized dynamics $|\Psi_d(mT/M)\rangle$, where $T$ is the final time, $M$ is the number of time steps, and $m$ is an integer, $m=0,1,2,\dots,M$, and discuss the overlap between these two dynamics at the final time, $|\langle\Psi(T)|\Psi_d(T)\rangle|$.

\paragraph{State-dependent error bound~\cite{Hatomura2022}.}
By using the unitary invariance and the subadditivity of the Fubini-Study angle, we find the following inequality
\begin{equation}
|\langle\Psi(T)|\Psi_d(T)\rangle|\ge\cos\left(\sum_{n=1}^M\mathcal{L}_n\right),\quad\text{for }\sum_{n=1}^M\mathcal{L}_n\le\frac{\pi}{2},
\label{Eq.sdepin}
\end{equation}
where
\begin{equation}
\begin{aligned}
\mathcal{L}_n=\arccos&|\langle\Psi(nT/M)|\hat{U}_d\bm{(}nT/M,(n-1)T/M\bm{)} \\
&\times[\hat{U}\bm{(}nT/M,(n-1)T/M\bm{)}]^\dag|\Psi(nT/M)\rangle|. 
\end{aligned}
\label{Eq.Ln}
\end{equation}
Here, $\hat{U}\bm{(}nT/M,(n-1)T/M\bm{)}$ and $\hat{U}_d\bm{(}nT/M,(n-1)T/M\bm{)}$ are time-evolution operators which transform $|\Psi\bm{(}(n-1)T/M\bm{)}\rangle$ and $|\Psi_d\bm{(}(n-1)T/M\bm{)}\rangle$ into $|\Psi(nT/M)\rangle$ and $|\Psi_d(nT/M)\rangle$, respectively. 
Note that the identical initial states, $|\Psi(0)\rangle=|\Psi_d(0)\rangle$, are assumed. \\

Now we assume that the Hamiltonian is given by $\hat{H}(t)=\sum_k\hat{H}_k(t)$ and consider the first-order Trotter decomposition for digitization. 
That is, the time-evolution operator for the digitized dynamics is given by $\hat{U}_d\bm{(}nT/M,(n-1)T/M\bm{)}=\prod_k\exp\bm{(}-\frac{i}{\hbar}\frac{T}{M}\hat{H}_k(nT/M)\bm{)}$.

\paragraph{Dominant errors~\cite{Hatomura2022}.}
When $T/M\ll1$ holds, we can apply the Taylor expansion to Eq.~(\ref{Eq.Ln}) and it gives
\begin{equation}
\begin{aligned}
&\mathcal{L}_n\approx\frac{T^2}{2\hbar^2M^2}|\langle\Psi(nT/M)|\hat{A}(nT/M)|\Psi(nT/M)\rangle|, \\
&\hat{A}(nT/M)=\sum_{k,l}[\hat{H}_k(nT/M),\hat{H}_l(nT/M)]. 
\end{aligned}
\label{Eq.error.normal}
\end{equation}
Here, we assume $\hat{H}(nT/M)\approx\hat{H}\bm{(}(n-1)T/M\bm{)}$ and discretize the time-evolution operator for the reference dynamics.

\paragraph{Infidelity.}
The state-dependent inequality for the overlap (\ref{Eq.sdepin}) gives inequality for the infidelity
\begin{equation}
\begin{aligned}
\sqrt{1-|\langle\Psi(T)|\Psi_d(T)\rangle|^2}&\le\sqrt{1-\cos^2\left(\sum_{n=1}^M\mathcal{L}_n\right)} \\
&\approx\sum_{n=1}^M\mathcal{L}_n,\quad\text{for small }\sum_{n=1}^M\mathcal{L}_n. 
\end{aligned}
\end{equation}
Since each $\mathcal{L}_n$ is $\mathcal{O}(M^{-2})$ as one can find in Eq.~(\ref{Eq.error.normal}), the infidelity is $\mathcal{O}(M^{-1})$. 
Note that this scaling of the digitization errors is identical with the well-known result~\cite{Lloyd1996}, but this inequality can reveal true scaling for specific decomposition as discussed below.

\subsection{Trotter decomposition in the dynamical-invariant basis}

Now we consider dynamics described by the single eigenvector of the dynamical invariant, $|\Psi(t)\rangle=e^{i\kappa_k(t)}|\phi_k(t)\rangle$, and the first-order Trotter decomposition which devide the Hamiltonian into the diagonal part and the off-diagonal part in the dynamical-invariant basis, i.e., we devide the Hamiltonian into the first term and the second term in Eq.~(\ref{Eq.ham.divide}). 
Then, we find that the dominant errors (\ref{Eq.error.normal}) vanish because $\hat{A}(nT/M)$ is off-diagonal in the dynamical-invariant basis and the state is given by the single eigenvector of the dynamical invariant $|\Psi(t)\rangle=e^{i\kappa_k(t)}|\phi_k(t)\rangle$.

\paragraph{Dominant errors.}
By considering higher-order expansion, we find that dominant errors of the first-order Trotter decomposition in the dynamical-invariant basis are given by
\begin{equation}
\begin{aligned}
&\mathcal{L}_n\approx\frac{T^3}{6\hbar^3M^3}|\langle\phi_k(nT/M)|\hat{B}(nT/M)|\phi_k(nT/M)\rangle|, \\
&\hat{B}(nT/M)=\bm{[}\hat{H}_\mathrm{nd}(nT/M),[\hat{H}_\mathrm{nd}(nT/M),\hat{H}_\mathrm{d}(nT/M)]\bm{]},
\end{aligned}
\label{Eq.errors.DI}
\end{equation}
where $\hat{H}_\mathrm{d}(nT/M)$ and $\hat{H}_\mathrm{nd}(nT/M)$ are the diagonal part and the off-diagonal part of the Hamiltonian, i.e., the first term and the second term in Eq.~(\ref{Eq.ham.divide}), respectively. 
That is, the infidelity scales as $\mathcal{O}(M^{-2})$ which is better than the conventional prediction $\mathcal{O}(M^{-1})$. 
Note that this result is a generalization of the result in Ref.~\cite{Hatomura2023}.

\subsection{Application to digitized counterdiabatic driving}

Finally, we show a useful example, i.e., digitized counterdiabatic driving~\cite{Hegade2021,Chandarana2022,Hegade2022,Hatomura2023}.

\paragraph{Counterdiabatic driving.}
In shortcuts to adiabaticity by counterdiabatic driving~\cite{Demirplak2003,Berry2009}, we consider the following Hamiltonian
\begin{equation}
\begin{aligned}
&\hat{H}(t)=\hat{H}_\mathrm{ref}(t)+\hat{H}_\mathrm{cd}(t), \\
&\hat{H}_\mathrm{ref}(t)=\sum_nE_n(t)|n(t)\rangle\langle n(t)|, \\
&\hat{H}_\mathrm{cd}(t)=i\hbar\sum_{\substack{m,n \\ (m\neq n)}}|m(t)\rangle\langle m(t)|\partial_tn(t)\rangle\langle n(t)|, 
\end{aligned}
\label{Eq.cdham}
\end{equation}
where $\hat{H}_\mathrm{ref}(t)$ is the reference Hamiltonian and $\hat{H}_\mathrm{cd}(t)$ is the counterdiabatic Hamiltonian. 
The counterdiabatic Hamiltonian cancels out diabatic changes and we can realize adiabatic time evolution of the reference Hamiltonian within arbitrary time. 
It is obvious from the derivation~\cite{Demirplak2003,Berry2009}, but we can also confirm this fact as follows.

\paragraph{Dynamical invariant in counterdiabatic driving.}
For the Hamiltonian (\ref{Eq.cdham}), the following Hermitian operator
\begin{equation}
\hat{F}(t)=\sum_nf_n|n(t)\rangle\langle n(t)|,
\end{equation}
is the dynamical invariant, where $f_n$ is an arbitrary constant. 
That is, the set of eigenvectors of the reference Hamiltonian $\{|n(t)\rangle\}$ is that of the dynamical invariant in this system. 
Then, Eq.~(\ref{Eq.LRstate}) gives the adiabatic state of the reference Hamiltonian, i.e., 
\begin{equation}
|\Psi_\mathrm{ad}(t)\rangle=\sum_nc_n(0)e^{-\frac{i}{\hbar}\int_0^tdt^\prime E_n(t^\prime)}e^{-\int_0^tdt^\prime\langle n(t^\prime)|\partial_{t^\prime}n(t^\prime)\rangle}|n(t)\rangle.
\end{equation}

\paragraph{Digitized counterdiabatic driving.}
In digitized counterdiabatic driving~\cite{Hegade2021,Chandarana2022,Hegade2022,Hatomura2023}, we devide the time-evolution operator with the Hamiltonian (\ref{Eq.cdham}) into that of the reference Hamiltonian and that of the counterdiabatic Hamiltonian by using the first-order Trotter decomposition. 
Since the reference Hamiltonian and the counterdiabatic Hamiltonian are the diagonal part and the off-diagonal part of the total Hamiltonian (\ref{Eq.cdham}), the dominant errors of this decomposition is given by Eq.~(\ref{Eq.errors.DI}). 
It means that the scaling of digitization errors is $\mathcal{O}(M^{-2})$ in terms of the infidelity.

\section{Summary}

In this paper, we considered the first-order Trotter decomposition in the dynamical-invariant basis. 
By using the state-depdendent inequality~\cite{Hatomura2022}, we found that the infidelity scales as $\mathcal{O}(M^{-2})$, whereas conventional approaches predict $\mathcal{O}(M^{-1})$. 
We also pointed out that this decomposition includes digitized counterdiabatic driving~\cite{Hegade2021,Chandarana2022,Hegade2022,Hatomura2023}.

For experimental implementation, further decomposition may be required. 
In such cases, we should use decomposition, whose precision is better than $\mathcal{O}(M^{-2})$, not to lose the advantage of the present decomposition.

In addition to digitized counterdiabatic driving, there are other possible applications. 
For example, there is a quantum computation approach based on the Lewis-Riesenfeld theory~\cite{Sarandy2011}. 
Moreover, there is another method of shortcuts to adiabaticity based on the Lewis-Riesenfeld theory, i.e., invariant-based inverse engineering~\cite{Chen2010}.

\end{document}